\documentclass[pre,aps,ams,twocolumn,showpacs]{revtex4}
\usepackage{epsfig}
\usepackage{graphicx}
\usepackage{textcomp}
\usepackage{amsmath,amssymb}
\begin{document}
\title{Population extinction in a time-modulated environment}
\author{Michael Assaf$^1$, Alex Kamenev$^2$, Baruch Meerson$^1$}
\affiliation{$^1$ Racah Institute of Physics, Hebrew University of
Jerusalem, Jerusalem 91904, Israel\\$^2$ Department of Physics,
University of Minnesota, Minneapolis, Minnesota 55455, USA}
\begin{abstract}
The extinction time of an isolated population can be exponentially reduced by
a periodic modulation of its environment.  We investigate this effect using, as an example,
a stochastic branching-annihilation process with a time-dependent
branching rate.
The population extinction is treated in eikonal approximation,
where it is described  as an instanton trajectory of a proper reaction
Hamiltonian. The modulation of the  environment perturbs
this trajectory and synchronizes it with the modulation phase.
We calculate the corresponding change in the action along the instanton
using perturbation techniques supported by numerical calculations. The
techniques include a first-order theory with respect to the modulation
amplitude, a second-order theory in the spirit of the Kapitsa pendulum
effect, and adiabatic theory valid for low modulation frequencies.

\end{abstract}
\maketitle

\section{Introduction}
Extinction  of species after maintaining a long-lived
quasi-stationary self-regulating population is a striking manifestation of large deviations coming from
intrinsic stochasticity of processes of  birth-death type.
Not surprisingly, the extinction phenomenon is at the center stage of
population biology and epidemiology \cite{population}. More recently, it has attracted
interest
in the context of cell biochemistry \cite{bio}.  As birth-death processes are usually
far from equilibrium, they are also of much interest to
physics \cite{Gardiner,vankampen}.    Birth-death processes often occur in a time-varying
environment, and the variations of the environment manifest themselves as an
explicit time dependence of the birth and/or death rates.
Elucidating different regimes of \textit{exponential} reduction of the mean time to
extinction (MTE) of species due to the time variation of the
process rates is both important and interesting \cite{assessment}. There is
a large body of work on
the effects of environmental \textit{noise} on the MTE
of birth-death systems. Early theoretical works on this
subject assumed that the environmental noise which modulates the rates
is delta-correlated  \cite{Leigh,Lande}.  More
recently, effects of finite correlation time of the noise have also
been addressed in numerical simulations \cite{Johst}. Not
surprisingly, the simulation results provide only a partial
understanding of the complex interplay between the nonlinear
kinetics and intrinsic stochasticity of the self-regulating population on the one
hand, and the magnitude and spectral/correlation properties of the environmental
noise on the other hand. To get more insight into this type of problems, we
will follow the recent work of Escudero and Rodr\'{\i}guez
\cite{escudero} and consider a much simpler model, where the time
variation of the environment, as manifested in the rate modulation,
is \textit{periodic} in time. Although completely ignoring the noise aspect,
this model enables one to investigate, in a controlled and often analytic way, different
frequency and amplitude regimes of response of a self-regulating birth-death system
to the rate modulations. Furthermore, not all environmental variations
are noisy: some of them (such as daily, monthly, and annual cycles) occur in an almost periodic fashion.

The main objective of this work is to calculate, in different regimes of parameters, the exponential reduction in the
MTE of birth-death systems due to a sinusoidal rate modulation.
These calculations are
intimately related to finding the ``optimal path to extinction", or instanton connection in birth-death systems \cite{dykman1,Kamenev1}. The instanton connection emerges in eikonal
approximation to the original master equation, and it
describes the most probable sequence of
events that brings the birth-death  system from its long-lived quasi-stationary state
to extinction. As a
prototypical example we will consider the stochastic
branching-annihilation process $A\to\hspace{-4.3mm}^{\lambda}\hspace{2mm} 2A$ and $2A
\to\hspace{-4.3mm}^{\mu}\hspace{2mm} \emptyset$. This
single-species birth-death process can be viewed as a simplified
version of the well-known Verhulst logistic model \cite{Nasell}. Its constant-rate
version was previously analyzed by a number of methods employing the large parameter
$\Omega\equiv\lambda/\mu\gg 1$ that corresponds to a large average number of individuals in the quasi-stationary state
at times short compared to the MTE \cite{Kamenev1,turner,kessler,Assaf1}.

The main theoretical tool that we employ in this work is an eikonal method which assumes that $\Omega \gg 1$. We apply the eikonal method to
the exact evolution equation for the probability
generating function that encodes the original continuous-time master equation \cite{Gardiner,vankampen}.
In the geometrical-optics order of the eikonal approximation, the evolution equation for the probability
generating function
reduces to a Hamilton-Jacobi equation, with an effective
Hamiltonian that explicitly depends on time.
We explore, analytically and numerically,  the phase trajectories generated by this
Hamiltonian and find the perturbed
instanton: a perturbed heteroclinic phase trajectory that corresponds to extinction. Calculating the action along the instanton yields
an approximation for the logarithm of the MTE \cite{dykman1,Kamenev1}. Because of the explicit
time dependence of the eikonal Hamiltonian, analytical progress is possible only via perturbation approaches. Therefore, we use the eikonal method in conjunction
with two different perturbation techniques that employ additional
small parameters.  The first of them is linear theory (LT): a theory based on linearization
of Hamilton equations with respect to the modulation amplitude. For birth-death processes,
the LT has been recently used  by Escudero and
Rodr\'{\i}guez \cite{escudero}. A closely related  theory was earlier developed by Dykman
\textit{et al.} \cite{dykman,dykmanreview}  in the context of large
fluctuations and escape in continuous (Langevin-type) stochastic systems driven by a
time-periodic signal. Dykman \textit{et al.}  showed that in LT
the modulation signal removes the degeneracy of the unperturbed instanton
trajectories with respect to time shift. This leads to the selection of the optimal instanton
which is
synchronized  with the phase of the modulation signal. In general,
the action along the optimal instanton is smaller than that the action along
the unperturbed instanton, leading to an exponential reduction of
the escape time  \cite{dykman,dykmanreview}.

By employing a theorem due to Melnikov \cite{Melnikov,guckenheimer}, Escudero and
Rodr\'{\i}guez \cite{escudero} proved the existence of a perturbed
instanton in the branching-annihilation process with \textit{weakly} modulated reaction rates.
We extend the LT by calculating analytically the action along
the instanton, which yields the logarithm of MTE.  We also show numerically that the perturbed instanton persists for \textit{any}
reasonable modulation amplitudes, and for all modulation frequencies. Furthermore, we show that the LT gives satisfactory results only when the modulation frequency $\omega$ is smaller than or comparable with the relaxation rate of the system $\lambda$. When $\omega \gg \lambda$, the first-order correction to action, as predicted by the LT, turns out to be
exponentially small with respect to the rescaled frequency $\omega/\lambda$.
It is the second-order correction in the modulation amplitude that becomes dominant in this regime.
We calculate this correction by employing (a Hamiltonian extension of) the
Kapitsa method, see \textit{e.g.} \cite{landau}.

The opposite, low-frequency regime is both the simplest and most important as, for the same modulation amplitude, the exponential reduction of the MTE turns out to be the largest here. This regime can be efficiently dealt
with in adiabatic approximation, for any reasonable modulation amplitude.  Here one assumes that
the extinction rate corresponding to the unperturbed problem is known, obtains the instantaneous extinction rate and averages it over the modulation period.  This simple procedure yields the average extinction rate, and hence the MTE, of the perturbed problem. If the knowledge of the extinction rate of the unperturbed problem incudes a pre-exponent, the adiabatic approximation yields a (modified) pre-exponent of the MTE of the perturbed problem: a significant improvement over the geometrical-optics order of the eikonal method.

The remainder of the paper is organized as follows. We begin
Section~\ref{model} with a brief introduction to the geometrical-optics  theory of
extinction in birth-death processes with \textit{time-independent}
rates using  the
branching-annihilation model as an example. In Section~\ref{theory} we
present the eikonal theory of population extinction for a time-periodic rate modulation.
We begin Section~\ref{theory} by developing the LT, and continue by  addressing the high-frequency limit $\omega \gg \lambda$
and employing the Kapitsa method. We conclude Section~\ref{theory} by reporting
numerical solutions of the Hamilton equations, in order to verify our theoretical results and extend them beyond the validity domain of the perturbation techniques. Section~\ref{adiabatic} deals with the low-frequency limit by means of adiabatic approximation. Theoretical results obtained in this limit are compared with numerical solutions of the master equation.  Section~\ref{summary} presents a brief summary and discussion of our results.

\section{Master equation, probability generating function, and geometrical optics of extinction}\label{model}
Here we present a brief introduction into an eikonal theory of population extinction in a time-independent environment \cite{dykman1,Kamenev1}, using the
prototypical example of the branching-annihilation process
$A\to\hspace{-4.3mm}^{\lambda}\hspace{2mm} 2A$ and $2A
\to\hspace{-4.3mm}^{\mu}\hspace{2mm} \emptyset$, where
$\lambda,\mu>0$ are the reaction rate constants \cite{turner,Kamenev1,kessler,Assaf1}.
We assume in this section that $\lambda$ and $\mu$ are independent of time.

At the level of deterministic modeling, the dynamics of the average
number of individuals $\bar{n}(t)$ is described by the
(mean-field) rate equation
\begin{equation}\label{rateeq}
\frac{d\bar{n}}{dt}=\lambda \bar{n}-\mu \bar{n}^2\,.
\end{equation}
This equation predicts an attracting fixed point
$\bar{n}=\lambda/\mu\equiv \Omega \gg 1$.  The rate equation ignores the
intrinsic noise that comes from the discreteness of individuals and
stochastic character of the reactions. This noise determines the
\textit{probability distribution} of the actual values of $n$ and, in particular,
a quasi-stationary distribution around the average value $\Omega \gg 1$ that sets in on
a relaxation time scale $\lambda^{-1}$, see below. On this time scale the mean-field picture
correctly describes
the average number of individuals in the stochastic process. At
sufficiently long times, however, the noise invalidates the mean-field picture
completely due to the existence, in the stochastic process, of the absorbing
state $n=0$. In other words, the stochastic process $A\to 2A$ and $2A
\to\emptyset$  eventually suffers a rare sequence of events
that brings the system into the empty state.

The intrinsic noise is accounted for quantitatively by the master
equation which describes the time evolution of ${\cal P}_n(t)$: the
probability to have $n$ individuals at time $t$. In the branching-annihilation example, the
continuous-time master equation, for $n \geq 1$, reads
\begin{eqnarray}
\label{master} \frac{d{\cal
P}_{n}(t)}{dt}&=&\frac{\mu}{2}\left[(n+2)(n+1)
{\cal P}_{n+2}(t)-n(n-1){\cal P}_{n}(t)\right]\nonumber\\
&+&\lambda\left[(n-1){\cal P}_{n-1}(t)-n{\cal P}_{n}(t)\right]\,.
\end{eqnarray}
The master equation can be conveniently recast with the help of
the probability generating function
\cite{Gardiner,vankampen}
\begin{equation}\label{genprobgen}
G(\wp,t)=\sum_{n=0}^{\infty} \wp^n {\cal P}_n(t)\,,
\end{equation}
where $\wp$ is an auxiliary variable. $G(\wp,t)$ encodes all the
probabilities ${\cal P}_n(t)$, as the latter ones are given by the coefficients of
the Taylor expansion of $G(\wp,t)$ around $\wp=0$. As the probabilities are normalizable to 1, $\sum_0^{\infty}{\cal P}_n(t)=1$, the
generating function satisfies the condition
\begin{equation}\label{norm}
    G(1,t)=1\,.
\end{equation}
The time-dependent
moments of the distribution ${\cal P}_n(t)$ can be expressed through the derivatives
of the generating function at $\wp=1$, \textit{e.g.} $\langle
n\rangle(t) \equiv \sum_n n {\cal P}_n(t) = \left.\partial_\wp
G(\wp,t)\right|_{\wp=1}$.

After a simple algebra,
Eq.~(\ref{master}) can be transformed into an
exact evolution equation for $G(\wp,t)$:
\begin{equation}\label{Gdot}
\frac{\partial G}{\partial t} =
\frac{\mu}{2}(1-\wp^2)\frac{\partial^2 G}{\partial \wp^2}+
\lambda\wp(\wp-1)\frac{\partial G}{\partial \wp}\,.
\end{equation}
The absorbing state at $n=0$ corresponds to the stationary solution $G_s(\wp)=1$. The presence of the absorbing state is manifested by the absence in Eq.~(\ref{Gdot}) of a term proportional to $G(\wp,t)$.

An initial value problem for Eq.~(\ref{Gdot}) can be solved by
expanding $G(\wp,t)$ in the eigenmodes of a Sturm-Liouville problem
related to the non-Hermitian operator
$$
\hat{{\cal H}}=\frac{\mu}{2}(1-\wp^2)\frac{\partial^2}{\partial \wp^2}+
\lambda\wp(\wp-1)\frac{\partial}{\partial \wp}\,,
$$
see Ref.~\cite{Assaf1} for detail. As in many other problems of
extinction of species, there is a zero eigenvalue $E_0=0$ which
describes the absorbing state, and an infinite discrete set of
eigenvalues $\{E_n\}_{n=1}^{\infty}$ which describe an exponential
decay with time of the rest of eigenmodes \cite{Assaf1,Assaf2}. For
$\Omega\gg 1$ the eigenvalue $E_1$  is exponentially small in $\Omega$, and much
less than the eigenvalues $E_2,E_3,\dots,$ which are of the order of the
relaxation rate of the system $\lambda$. Therefore, there are two
widely different time scales in the problem.  The decay, on the time
scale $\lambda^{-1}$, of the higher eigenmodes corresponds to a
rapid relaxation of the probability distribution ${\cal P}_n(t)$ to
the quasi-stationary distribution, 
peaked at $n\simeq\Omega$. The much slower
decay, on the time scale $E_1^{-1}$, of the lowest excited eigenmode
[accompanied by a slow growth of the extinction probability
${\cal P}_0(t)$] corresponds to an exponentially slow decay of the
quasi-stationary distribution and to extinction of the species. A
simple approximation to the eigenvalue $E_1$ is provided by geometrical optics, by which we mean 
the leading order of the eikonal approximation \cite{Kamenev1}. Although it gives only exponential accuracy, the geometric optics
can be readily applied to
systems of two species \cite{KM,dykmanSIS} and to time-dependent problems
\cite{escudero,dykmanSIS,AKM}: the focus of the present work.

Employing the eikonal ansatz $G(\wp,t)=\exp[-S(\wp,t)]$ in
Eq.~(\ref{Gdot}) and neglecting $\partial ^2 S/\partial\wp^2$, we
arrive at a Hamilton-Jacobi equation for $S(\wp,t)$ in the
$\wp$-representation:
\begin{eqnarray} \label{HJeq}
\frac{\partial S}{\partial
t}+\frac{\mu}{2}(1-\wp^2)\left(\frac{\partial S}{\partial
\wp}\right)^2-\lambda\wp(\wp-1)\frac{\partial S}{\partial \wp}=0\,.
\end{eqnarray}
Introducing a canonically conjugate coordinate $q=-\partial
S/\partial \wp$ and shifting the momentum $p=\wp-1$, we arrive at a
one-dimensional Hamiltonian flow, where $p$ plays the role of the
momentum  \cite{Kamenev1}:
\begin{equation}\label{H1}
H(q,p)=\left[\lambda(1+p)-\frac{\mu}{2}(2+p)q\right]\, qp\,.
\end{equation}
The Hamilton equations are
\begin{eqnarray}
\dot{q} &=& \frac{\partial H}{\partial p} = q \left[\lambda(1+2p)-\mu (1+p)q\right] \,, \label{qdot} \\
\dot{p} &=& -\frac{\partial H}{\partial q} = p
\left[\mu(2+p)q-\lambda(1+p)\right]\label{pdot}\,.
\end{eqnarray}
\begin{figure}
\includegraphics[width=6.0cm,height=5.5cm,clip=]{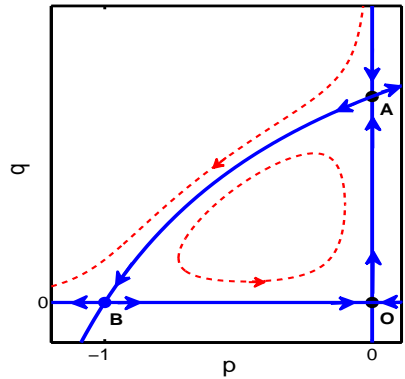}
\caption{(Color online.) The phase plane of the unperturbed Hamiltonian (\ref{H1}).
The solid lines denote zero-energy trajectories.} \label{phaseplane}
\end{figure}
As the geometrical-optics Hamiltonian (\ref{H1}) does not depend explicitly on time, it is
conserved: $H(q,p)=E=const$, and the problem is
integrable. The phase plane $(q,p)$, defined by this
Hamiltonian flow, provides a useful visualization of the
dynamics, see Fig.~\ref{phaseplane}.
The attracting fixed point $\bar{n}=\Omega$ and the repelling fixed
point $\bar{n}=0$ of the one-dimensional rate
equation (\ref{rateeq}) become hyperbolic points $A(q=\Omega,\,p=0)$ and
$O(q=0,p=0)$ of the phase plane $(q,p)$. They belong to the mean-field zero-energy line $p=0$.
[The equality $H(q,0)=0$ reflects the probability
conservation.] Another zero-energy line is the extinction line $q=0$. Of special importance, however, is the
non-trivial zero-energy line
\begin{equation}\label{q(p)}
q= q_0(p)=\frac{2\Omega(1+p)}{(2+p)}\,.
\end{equation}
This line, which includes still another hyperbolic fixed point $B(q=0,\,p=-1)$
(we call it the fluctuational point), gives a geometrical-optics description of the
quasi-stationary probability distribution. The $-1\leq p\leq 0$
segment of this line is a heteroclinic trajectory. It exits, at
time $t=-\infty$, the hyperbolic fixed point $A$ along its
unstable manifold and enters, at time $t=\infty$, the hyperbolic
fixed point $B(q=0,\,p=-1)$ along its stable manifold. The heteroclinic
trajectory (often referred to as the optimal path to extinction, or
the instanton \cite{dykman2,freidlin,graham}) describes
the most probable sequence of events bringing the system
from the quasi-stationary state to extinction.  Up
to a pre-exponent, that is missed by the geometrical-optics order of the eikonal theory,
the MTE can be approximated as $\tau_{ex} \sim
\exp({\cal S}_0)$, where
\begin{equation}\label{action1}
{\cal S}_0 = -\int_0^{-1}
q_0(p)\,dp=2\Omega(1-\ln 2)\,.
\end{equation}
The time-dependent instanton coordinate $q_0(t-t_0)$ and momentum $p_0(t-t_0)$ are the following:
\begin{equation}\label{unpert}
q_0(t-t_0)=\frac{2\Omega}{2+e^{\lambda(t-t_0)}}\;,\;\;\;p_0(t-t_0)=-\frac{1}{1+e^{-\lambda(t-t_0)}}\,,
\end{equation}
where $t_0$ is an arbitrary time shift.

The eikonal approximation is valid when ${\cal S}_0 \gg 1$, that is $\Omega \gg 1$. Equation (\ref{action1}),
obtained by Elgart and Kamenev \cite{Kamenev1}, coincides with the leading order term of the logarithm of the MTE found by other methods (also employing the strong inequality $\Omega \gg 1$), by Turner and Malek-Mansour
\cite{turner}, Kessler and Shnerb \cite{kessler}, and Assaf
and Meerson \cite{Assaf1}.

\section{Extinction in a time-varying environment: perturbation techniques for the instanton}\label{theory}

Now let the branching rate $\lambda$ in the master equation
(\ref{master}), and in the evolution equation (\ref{Gdot}) for $G(\wp,t)$, vary sinusoidally in time:
\begin{equation}\label{rates}
\lambda(t)=\lambda_0\left(1+\varepsilon \cos \omega t\right)\,,
\end{equation}
where $|\varepsilon|<1$, and $\varepsilon$ and $\omega$ are the modulation amplitude and frequency, respectively.
The eikonal approximation now yields the following
\textit{time-dependent} Hamiltonian:
\begin{equation}\label{pertHam}
    H(q,p,t)=H_0(q,p) + \varepsilon H_1(q,p,t) \,,
\end{equation}
where
\begin{equation}\label{h0pert}
H_0(q,p)=\lambda_0(1+p)pq-\frac{\mu_0}{2}(2+p)p q^2\,,
\end{equation}
and
\begin{equation}\label{h1pert}
H_1(q,p,t)=\lambda_0(1+p)pq \,\cos \omega t\,.
\end{equation}
For the unperturbed Hamiltonian $H_0(q,p)$, the instanton (\ref{q(p)}) and
the corresponding action ${\cal S}_0$ along it, Eq.~(\ref{action1}), are known.
Our strategy will
be to calculate the action along the \textit{perturbed} instanton:
the instanton of the perturbed, time-dependent Hamiltonian (\ref{pertHam}).

Following Escudero and Rodr\'{\i}guez \cite{escudero},   let us
consider the Poincar\'{e} map
\textit{P}$_{\varepsilon}^{t_0}=\{q,p,t|t=t_0\in[0,T]\}$: the
projection of three-dimensional trajectories of the non-autonomous
system (\ref{pertHam}) in the $(q,p,t)$ space on the $(q,p)$ plane
at section time $t_0\in[0,2\pi/\omega]$. In the perturbed system the
hyperbolic points $A$ and $B$ are also perturbed. We will denote the
perturbed fixed points by $A_{\varepsilon}^{t_0}$ and
$B_{\varepsilon}^{t_0}$, respectively. For generic Hamiltonians the
existence of the perturbed hyperbolic points is guaranteed, by the
Poincar\'{e}-Birkhoff fixed point theorem \cite{guckenheimer}, only
for sufficiently small $\varepsilon$. For the birth-death
Hamiltonian we are dealing with here the perturbed fixed points
exist for \textit{any} $|\varepsilon|<1$.  The point
$B_{\varepsilon}^{t_0}=(q=0,p=-1)$ remains unchanged \cite{badeg},
whereas the point $A_{\varepsilon}^{t_0}$  can be easily found from
the time-dependent rate equation (\ref{qdot}) with $p=0$ and
$\lambda$ from Eq.~(\ref{rates}):
\begin{equation}
\label{dotqmf}
\dot{q}=q \left[\lambda_0 (1+\varepsilon \cos \omega t)-\mu q \right]\,.
\end{equation}
The solution of this equation, for an arbitrary initial condition $q(0)$, is the following:
\begin{equation}\label{qt}
q(t)=\frac{q(0)\exp\left(\lambda_0
t+\frac{\varepsilon\lambda_0\sin\omega
t}{\omega}\right)}{1+\frac{\mu_0}{\lambda_0}q(0)\int_{0}^{\lambda_0
t}\exp\left[s+\frac{\varepsilon\lambda_0\sin(\omega
s/\lambda_0)}{\omega}\right]ds}\,,
\end{equation}
see also Ref.~\cite{escudero}. At long times this solution becomes periodic in time, and can be written as
\begin{equation}
\label{periodicA}
q(t)=\frac{\lambda_0}{\mu_0}\frac{\exp\left(\frac{\varepsilon\lambda_0\sin\omega
t}{\omega}\right)}{\displaystyle\sum_{n=-\infty}^{\infty}\frac{I_n\left(\frac{\varepsilon\lambda_0}
{\omega}\right)\,\cos(n\omega t-\phi_n-n\pi/2)}{\sqrt{1+n^2
\omega^2/\lambda_0^2}}}\,,
\end{equation}
where $I_n(\dots)$ is the modified Bessel function, and
$\phi_n=\arctan (n\omega/\lambda_0)$. Putting in Eq.~(\ref{periodicA}) $t=t_0$ we obtain
$q$ of the perturbed fixed point $A_{\varepsilon}^{t_0}$, whereas its $p$ remains zero.

Existence of the perturbed fixed points is a necessary but, in general,
insufficient condition for the existence of a heteroclinic trajectory connecting the unstable manifold of
$A_{\varepsilon}^{t_0}$ and the stable manifold of
$B_{\varepsilon}^{t_0}$. In general, one should also establish the existence, in the perturbed problem, of
the unstable and stable manifolds themselves, and of their intersection \cite{guckenheimer,escudero}.
We verified analytically and numerically, see below, that the unstable manifold of $A_{\varepsilon}^{t_0}$ and the
stable manifold of $B_{\varepsilon}^{t_0}$ do intersect. Let us denote the perturbed instanton connection by the pair $q(t,t_0),p(t,t_0)$. As the energy is not conserved, the
action along the perturbed instanton now includes an integral of $H$ over time:
\begin{eqnarray}
{\cal S}&=&  \int_{-\infty}^{\infty}\!
\left\{p(t,t_0)\dot{q}(t,t_0)-H_0\left[q(t,t_0),p(t,t_0)\right]\right.\nonumber\\
  &-&\left.\varepsilon H_1\left[q(t,t_0),p(t,t_0),t\right]\right\} dt\,, \label{action}
\end{eqnarray}
where $\dot{q}(t,t_0) = dq/dt$, and the unperturbed Hamiltonian $H_0(q,p)$ is invariant to the
specific choice of $t_0$.  Escudero and Rodr\'{\i}guez \cite{escudero} applied the Melnikov theorem \cite{guckenheimer,Melnikov} and proved that a perturbed instanton
exists in this problem for sufficiently small $\varepsilon$. To this end they calculated (an approximation for)
the distance between the unstable and stable
manifolds, which is given by the Melnikov function, see below, and showed that it vanishes for a
specific choice of $t_0$. We briefly reproduce their derivation in subsection \ref{linear}, along with new results: a linear-theory calculation of the action (\ref{action}) and the corresponding reduction in the logarithm of the MTE. We also show, in subsection \ref{Kapitsa}, that the perturbed instanton
exists for high-frequency perturbations,
$\omega\gg\lambda_0$ at any $|\varepsilon|<1$.  Furthermore, we report, in subsection
\ref{numerics}, strong numerical evidence that the perturbed instanton exists for any $|\varepsilon|<1$ and any $\omega$.

\subsection{Linear correction to action}\label{linear}
Here we consider the linear theory (LT) which assumes that the term
$\varepsilon H_1(q,p,t)$ in Eq.~(\ref{pertHam}) can be treated
perturbatively. For this assumption to hold it is sufficient to
demand the strong inequality $|\varepsilon|\ll 1$. As we will see in
subsection \ref{Kapitsa}, at high frequencies, $\omega\gg\lambda_0$,
the strong inequality $|\varepsilon|\ll 1$ becomes unnecessary. For
completeness, we will derive Eqs. (\ref{S1}) and (\ref{Melnikov}) in
a general form, before dealing with the particular example of
branching and annihilation.  In the first order in $\varepsilon$ the
perturbed instanton of $H(q,p,t)$ is described by the equations
\begin{eqnarray}
q(t,t_0) &=& q_0(t-t_0)+\varepsilon q_1(t,t_0) \,, \nonumber\\
p(t,t_0) &=& p_0(t-t_0)+\varepsilon p_1(t,t_0) \,,
\end{eqnarray}
where $q_0(t-t_0)$ and $p_0(t-t_0)$ stand for the (known) instanton solution of the unperturbed
equations
\begin{eqnarray}
\dot{q}_0 &=& \frac{\partial H_0(q_0,p_0)}{\partial p_0} \,,
\nonumber\\ \dot{p}_0 &=& - \frac{\partial H_0(q_0,p_0)}{\partial
q_0} \,, \label{pq0}
\end{eqnarray}
that is Eqs.~(\ref{qdot}) and (\ref{pdot}) with $\lambda=\lambda_0$.

To calculate the action (\ref{action}) we expand the integrand in
$\varepsilon$ and obtain, in the first order,
\begin{eqnarray}
 &(&\hspace{-2mm}p_0+\varepsilon p_1) (\dot{q}_0+\varepsilon \dot{q}_1)
 -H_0(q_0,p_0)
- \varepsilon q_1 \frac{\partial H_0(q_0,p_0)}{\partial q_0}\nonumber\\
&-&\varepsilon p_1\frac{\partial H_0(q_0,p_0)}{\partial p_0}
-\varepsilon H_1(q_0,p_0,t)\simeq p_0\dot{q}_0-H_0(q_0,p_0)\nonumber\\
 &+&\varepsilon p_1\dot{q}_0+\varepsilon p_0\dot{q}_1
 + \varepsilon q_1\dot{p}_0-\varepsilon p_1\dot{q}_0 -\varepsilon H_1(q_0,p_0,t)\,.
\end{eqnarray}
After the integration the first two terms yield the
unperturbed action ${\cal S}_0$. The third term cancels out with the
sixth term, while the fourth term cancels out, after integration by
parts, with the fifth term. The result is ${\cal S}(t_0) = {\cal S}_0 +\Delta {\cal S}(t_0)$, where
\begin{equation}
\Delta {\cal S}(t_0)=-\varepsilon
\int_{-\infty}^{\infty}H_1\left[q_0(t-t_0),p_0(t-t_0),t\right]\,
dt\,. \label{S1}
\end{equation}
Recall that the integration is performed along the unperturbed instanton. To find the \textit{optimal} correction to action we have to \textit{minimize} ${\cal S}(t_0)$ with respect to $t_0$ (compare to Refs. \cite{dykman,dykmanreview}). This yields the following equation for $t_0$:
\begin{eqnarray}
\frac{d{\cal S}(t_0)}{dt_0} &=& \varepsilon
\int_{-\infty}^{\infty}\left(\frac{\partial H_1}{\partial q_0}
\dot{q}_0+\frac{\partial H_1}{\partial
p_0}\dot{p}_0\right)\,dt\nonumber\\ & \equiv & - \varepsilon
\int_{-\infty}^{\infty} \left\{ H_0, H_1 \right\}_0\,dt=0\,,
\label{Melnikov}
\end{eqnarray}
where $\{H_0,H_1\}_0$ is the Poisson bracket evaluated on the unperturbed instanton.
The quantity $M(t_0)=\int_{-\infty}^{\infty} \left\{ H_0, H_1 \right\}_0\,dt$
is the Melnikov function  of the perturbed problem
\cite{escudero,Melnikov,guckenheimer}. It is proportional to the
distance between the unstable and stable manifolds of
$A_{\varepsilon}^{t_0}$ and $B_{\varepsilon}^{t_0}$, respectively
\cite{guckenheimer}. That $M(t_0)$ has simple zeros yields a sufficient
condition for the existence of the perturbed
instanton. These zeros are the critical points of the function
${\cal S}(t_0)$. By finding the critical value of $t_0$ for which
${\cal S}(t_0)$ has its minimum, we obtain the minimal
action along the instanton of the perturbed problem
(\ref{S1}). It is the $\min \Delta {\cal S}(t_0)$ that yields the
reduction of the logarithm of
the MTE, caused by the time-dependence of the rate.   Note that
when $H_0$ commutes with $H_1$, $M(t_0)$ vanishes
identically at any $t_0$, but this degenerate case is hardly of interest.

The validity of the linear eikonal theory is determined by the double strong inequality
\begin{equation}\label{double}
1 \ll \Delta{\cal S} \ll {\cal S}_0\,.
\end{equation}
The left inequality is
required for the eikonal approximation to be valid, while the right inequality
is needed for the linearization to be valid. Note that  the rate
variation does not have to be time-periodic for the
LT to hold: it suffices to demand that the rate variation be bounded
by a sufficiently small value.

\begin{figure}
\includegraphics[width=6.0cm,height=5.5cm,clip=]{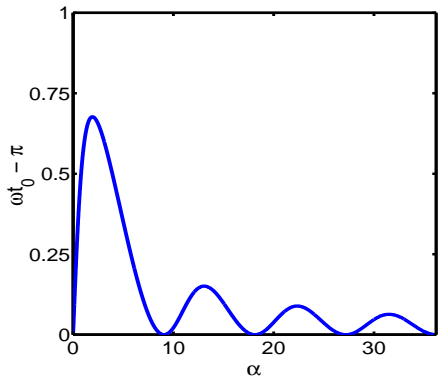}
\caption{(Color online.) The difference $\omega t_0-\pi$, see Eq.~(\ref{t1res}), as
a function of $\alpha=\omega/\lambda_0$.} \label{t0}
\end{figure}

Now we return to the branching-annihilation example.
In the first order in $\varepsilon$, the perturbed fixed point $A_{\varepsilon}^{t_0}$ is
determined by the equation
\begin{equation}\label{perturbed_A}
\frac{q_A}{\Omega}= 1+\frac{\varepsilon \lambda_0 \left( \lambda_0 \cos \omega t+ \omega \sin\omega t\right)}{\omega^2+\lambda_0^2}
\end{equation}
evaluated at $t=t_0$.  Substituting Eqs.~(\ref{unpert}) and (\ref{h1pert}) in Eq.~(\ref{S1}), we obtain
\begin{equation}
\Delta{\cal S}(t_0) = 2\varepsilon\Omega\int_{-\infty}^{\infty}
\frac{\cos(\tau/\lambda_0+t_0)e^{\tau}}{(1+e^{\tau})^2(2+e^{\tau})}d\tau\,.
\label{precalc}
\end{equation}
Calculating the integral we arrive at
\begin{eqnarray}
&&\frac{{\cal S}(t_0)}{{\cal S}_0}=
1+\frac{\pi\varepsilon}{(1-\ln 2)\sinh(\pi\omega/\lambda_0)}\nonumber\\
&&\times \left\{\frac{\omega}{\lambda_0} \cos \omega
t_0 -\sin\left[\omega\left(\frac{\ln
2}{\lambda_0}\!+\!t_0\right)\right]+\sin \omega
t_0 \right\},\nonumber\\
\label{s1calc}
\end{eqnarray}
where ${\cal S}_0$ is given by Eq.~(\ref{action1}). The minimal action is ${\cal
S}(t_0^*)$, where $t_0^*$ is the solution of the trigonometric equation for $t_0$:
\begin{eqnarray}
&&\hspace{-7mm}\frac{d{\cal S}(t_0)}{dt_0}\equiv - \varepsilon M(t_0)=-\frac{2\pi\varepsilon
\Omega\omega}{\sinh(\pi\omega/\lambda_0)}\nonumber\\
&\times&\left\{\frac{\omega}{\lambda_0} \sin\omega t_0+\cos\left[\omega\left(\frac{\ln
2}{\lambda_0}\!+\!t_0\right)\right]-\cos\omega t_0\right\}=0,\nonumber\\ \label{derresult}
\end{eqnarray}
subject to the condition $d^2{\cal S}(t_0)/dt_0^2>0$. The solution
is
\begin{equation}\label{t1res}
\omega t_0^*=\pi+\arctan\left[\frac{1-\cos(\alpha\ln 2)}{\alpha-\sin(\alpha\ln 2)}\right]\,,
\end{equation}
where $\alpha=\omega/\lambda_0$, and we assume here and in the
remainder of this subsection that $\varepsilon>0$ \cite{escudero1}.
Equation (\ref{t1res}) shows that the optimal instanton becomes
synchronized with the rate modulation, similarly to what happens in
the context of escape in Langevin-type stochastic systems driven by
a time-periodic signal \cite{dykman,dykmanreview}. Figure~\ref{t0}
depicts the difference $\omega t_0-\pi$ as a function of $\alpha$.
This difference is small at $\alpha \ll 1$, $\omega t_0-\pi \simeq
(\ln 2)^2/[2(1-\ln 2)]\alpha$, and at  $\alpha\gg 1$: $\omega
t_0-\pi \simeq\alpha^{-1}[1-\cos(\alpha\ln 2)]$. It vanishes for
$\alpha=2\pi n/\ln 2$, where $n=1,2,\dots$. For $\alpha \sim 1$ the
difference $\omega t_0-\pi$ is of the order of one.

\begin{figure}
\includegraphics[width=7.0cm,height=6cm,clip=]{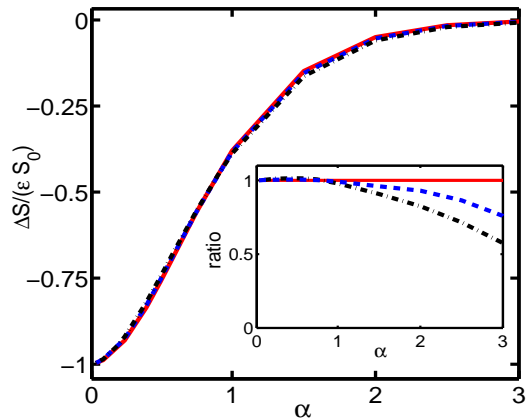}
\caption{(Color online.) The correction to action $\Delta{\cal S}$,
rescaled by $\varepsilon {\cal S}_0$, versus the rescaled modulation
frequency $\alpha=\omega/\lambda_0$. Dashed and dash-dotted lines:
predictions of the linear theory, $\Delta{\cal S}_{LT}$, see
Eq.~(\ref{minact}), for $\varepsilon=0.04$ and $\varepsilon=0.1$,
respectively. Solid line: $\Delta{\cal S}_{num}$ found by solving
the Hamilton equations numerically, see subsection \ref{numerics}.
On this scale the three lines are indistinguishable. The inset shows
the ratios of $\Delta{\cal S}_{LT}$ and $\Delta{\cal S}_{num}$ for
$\varepsilon=0.04$ (the dashed line), and $\varepsilon=0.1$ (the
dash-dotted line). The disagreement at high frequencies demands an
account of what we call the Kapitsa correction, calculated in
Section \ref{Kapitsa}.} \label{extvsw}
\end{figure}

Now we use Eqs.~(\ref{s1calc}) and (\ref{t1res}) to find the perturbed action:
\begin{eqnarray}
\frac{{\cal S}_{min}}{{\cal S}_0}&=&1+\frac{\Delta{\cal S}}{{\cal S}_0}=1-\frac{\pi\varepsilon}{(1-\ln 2)\sinh(\pi\alpha)}\nonumber\\
&\times&\left\{\left[\sin(\alpha\ln
2)-\alpha\right]^2+\left[\cos(\alpha\ln
2)-1\right]^2\right\}^{1/2}\,. \nonumber\\\label{minact}
\end{eqnarray}
Figure~\ref{extvsw} depicts
$\Delta{\cal S}/(\varepsilon \Omega)$ as a function of $\alpha$ as predicted by Eq.~(\ref{minact}). The maximum
effect of the rate modulation is obtained at
$\alpha \to 0$. Recall that the perturbed action yields (in the leading order in $\Omega$) the natural logarithm of the MTE of the
perturbed branching-annihilation problem.

The perturbative eikonal result is valid as long as the double
inequality (\ref{double}) is obeyed. The left inequality breaks down
at very small $\varepsilon$, whereas the right inequality breaks
down at not small $\varepsilon$ where linearization becomes invalid.
One can define, in analogy with the problem of activated escape in
Langevin-type stochastic systems \cite{dykman}, the logarithmic
susceptibility $\chi=\partial(\Delta{\cal
S})/\partial|\varepsilon|$. Within the framework of the LT [that is,
for intermediate values of the modulation amplitude that satisfy the
double inequality (\ref{double})], $\chi$ is independent of
$\varepsilon$:
\begin{eqnarray}\label{ls}
\chi&=&-\frac{2\pi\Omega}{\sinh(\pi\alpha)}\nonumber\\
&\times&\left\{\left[\sin(\alpha\ln
2)\!-\!\alpha\right]^2+\left[\cos(\alpha\ln
2)\!-\!1\right]^2\right\}^{1/2}\,.
\end{eqnarray}
At small and large modulation amplitudes $\chi$ becomes amplitude-dependent.

In the low-frequency
limit, $\alpha \ll 1$, one obtains from Eq.~(\ref{minact}):
\begin{equation}
\frac{\Delta{\cal S}}{\varepsilon{\cal
S}_0}=- 1+\frac{\alpha^2}{6}\left[\pi^2-\frac{(\ln
2)^3}{(1\!-\!\ln 2)^2}+\frac{(\ln 2)^4}{4(1\!-\!\ln
2)^2}\right]\,,\label{minactadia}
\end{equation}
where we have kept the leading and subleading terms. The leading term
coincides with the prediction (\ref{specadia}) from the adiabatic approximation,
see Section \ref{adiabatic}. In this limit the LT is valid for
$\Omega^{-1}\ll\varepsilon\ll 1$. The same criterion
applies for $\alpha \sim 1$, as
$\Delta{\cal S}/(\varepsilon\Omega)$ in Eq.~(\ref{minact}) is of
order $1$ there. In the high-frequency limit, $\alpha \gg 1$, the
correction $\Delta{\cal S}$ is exponentially small:
\begin{equation}\label{minactfast}
\frac{\Delta{\cal S}}{\varepsilon{\cal
S}_0}=-\frac{2 \pi \alpha\,e^{-\pi\alpha}}{1-\ln 2} \,.
\end{equation}
Here $\varepsilon$ need not be small compared to 1 to satisfy the
right inequality in Eq.~(\ref{double}).  Note that
Eqs.~(\ref{minact})-(\ref{minactfast}) with $\varepsilon$ replaced
by $|\varepsilon|$ are valid also for $\varepsilon<0$.

In Fig.~\ref{extvsw} Eq.~(\ref{minact}) is compared with the
correction to action found by solving the Hamilton equations
numerically (see subsection \ref{numerics} for details of the
numerics). One can see a good agreement at small and intermediate
frequencies, and a disagreement at high frequencies. What is the
reason for the disagreement? That the correction to action $\Delta
{\cal S}$, as predicted by the LT, becomes exponentially small at
large frequencies indicates that the $\varepsilon^2$-correction,
neglected by the LT, becomes dominant there. This conjecture is
validated in the next subsection, where we consider the
high-frequency case and calculate the $\varepsilon^2$-correction
using (a Hamiltonian extension of) the Kapitsa method.

\subsection{Quadratic correction to action: the Kapitsa correction \label{Kapitsa}}

Here we consider the high-frequency limit, $\omega \gg \lambda_0$
and calculate the correction to action which is of the second
order in $\varepsilon$. To this end we use (a Hamiltonian extension
of) the method that was developed long ago in the context of the
``Kapitsa pendulum": a rigid pendulum with a rapidly vibrating
pivot, see \textit{e.g.} \cite{landau}.
The Kapitsa method involves, as a first step, calculation of small
high-frequency corrections to the unperturbed coordinate and
momentum of the system. Because of the high frequency of the
perturbation these corrections are small even if $\varepsilon$ is
of order $1$. Using the high-frequency corrections, we construct a
canonical transformation which, by means of time averaging (that is, by
rectifying the high-frequency component of the motion), transforms
the original time-dependent Hamiltonian into an effective
\textit{time-independent} one. The effective Hamiltonian includes a
correction coming from the rectified high-frequency perturbation.
Finally, we find the perturbed instanton, emerging from this effective time-independent Hamiltonian,
and the action along the instanton.

The starting point of the derivation is the same Hamiltonian (\ref{pertHam}) with rates
given by Eq.~(\ref{rates}). We represent $q$ and $p$ as follows:
\begin{equation}\label{qtpt}
q(t)=X(t)+\xi(t)\,,\;\;\;p(t)=Y(t)+\eta(t)\,,
\end{equation}
where  $X$ and $Y$ are
slow variables, while $\xi$ and $\eta$ are small and rapidly oscillating. Now we expand the
Hamiltonian $H(q,p,t)$ around $q=X$ and $p=Y$ up to the second order
in $\xi$ and $\eta$:
\begin{eqnarray}
&&H(q,p,t)\simeq  H(X,Y,t)+\xi \frac{\partial
H(X,Y)}{\partial X}+\eta
\frac{\partial H(X,Y)}{\partial Y}\nonumber\\
&+&\frac{\xi^2}{2} \frac{\partial^2 H(X,Y)}{\partial
X^2}+\frac{\eta^2}{2} \frac{\partial^2 H(X,Y)}{\partial
Y^2}+\xi\eta\frac{\partial^2 H(X,Y)}{\partial X\partial
Y}\nonumber\\
&\equiv& \tilde{H}(X,Y,t)\,.\label{hamKapitsa}
\end{eqnarray}
The Hamilton equations become
\begin{equation}\label{pqdot}
\dot{q}=\dot{X}+\dot{\xi}\simeq\frac{\partial \tilde{H}(X,Y,t)}{\partial
Y}\,,\;\;\;\dot{p}=\dot{Y}+\dot{\eta}\simeq-\frac{\partial
\tilde{H}(X,Y,t)}{\partial X}\,.
\end{equation}
Now we
demand that the rapidly oscillating terms in Eqs.~(\ref{pqdot}) balance each other. This yields
\begin{equation}\label{xietad}
\dot{\xi}\simeq \varepsilon\lambda_0 X(2Y+1)\cos \omega
t\,,\;\;\dot{\eta}\simeq -\varepsilon\lambda_0 Y(Y+1)\cos \omega
t\,,
\end{equation}
where terms of the order of $\xi$ and $\eta$ have been neglected, but their time derivatives
(which are proportional to $\omega$ and
therefore large) have been kept. Treating $X$ and $Y$ as constants during the period of rapid oscillations $2 \pi/\omega$, we can easily solve Eqs.~(\ref{xietad}) and
obtain
\begin{equation}\label{xieta}
\xi(t)\simeq \frac{\varepsilon\lambda_0}{\omega} X(2Y+1)\sin \omega
t\,,\;\;\eta(t)\simeq -\frac{\varepsilon\lambda_0}{\omega}
Y(Y+1)\sin \omega t\,.
\end{equation}
Now it is clear that this perturbation scheme demands $|\varepsilon|
\lambda_0/\omega \ll 1$. As we have assumed $\omega \gg \lambda_0$,
$\varepsilon$ need not be small.

\begin{figure}
\includegraphics[width=6.75cm,height=5.75cm,clip=]{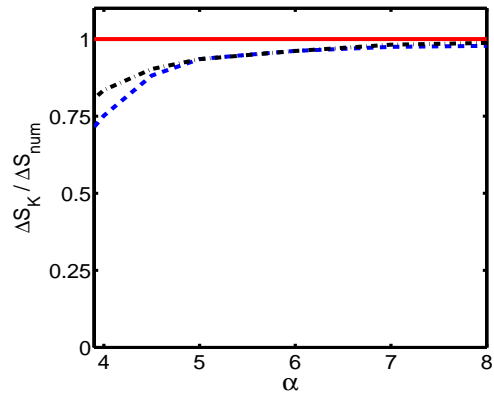}
\caption{(Color online.) The ratio of the Kapitsa correction $\Delta {\cal
S}_K$, Eq.~(\ref{Kapitsadelta}), and $\Delta{\cal S}_{num}={\cal
S}_{num}-2\Omega(1-\ln 2)$ for $\varepsilon=0.1$ (dashed line) and
$\varepsilon=0.2$ (dash-dotted line), versus
$\alpha=\omega/\lambda_0$. The agreement improves with an increase of $\alpha$ and $\varepsilon$.} \label{highfreq}
\end{figure}

Using Eqs.~(\ref{qtpt}) and (\ref{xieta}), we perform an almost canonical
transformation from the old variables $q$ and $p$ to the new variables $X$ and $Y$:
\begin{eqnarray}
&&\hspace{-9mm}q(X,Y,t)=\frac{X}{1-\frac{\varepsilon\lambda_0}{\omega}(2Y+1)\sin
\omega t}\nonumber\\
&&\hspace{-9mm}\simeq
X\left[1+\frac{\varepsilon\lambda_0}{\omega}(2Y\!+\!1)\sin \omega
t+\frac{\varepsilon^2\lambda_0^2}{\omega^2}(2Y\!+\!1)^2\sin^2\omega
t\right]\,,\nonumber\\
&&\hspace{-9mm}p(X,Y,t)=Y\left[1-\frac{\varepsilon\lambda_0}{\omega}(1+Y)\sin
\omega
t\right]\!.
\label{cantrans}
\end{eqnarray}
This transformation is canonical up to third order of
${\cal O}[(\lambda_0/\omega)^3] \ll 1$, as the Poisson brackets $\{q,p\}_{(X,Y)}=1+{\cal O}[(\lambda_0/\omega)^3]$. The
generating function of this transformation is
$$
F_2(q,Y,t)=q
Y\left[1-(\varepsilon\lambda_0/\omega)(Y+1)\sin\omega t\right]\,,
$$
see \textit{e.g.} Ref.~\cite{landau}. Now we transform to the new variables $X$ and $Y$,
$H^{\prime}= H+\partial F_2/\partial t$,
and average the new Hamiltonian $H^{\prime}$ over the period of rapid  oscillations $2 \pi/\omega$. This yields an effective Hamiltonian
\begin{eqnarray}
&&\hspace{-5mm}\bar{H}(X,Y)=\lambda_0 XY(1+Y)-\frac{\mu_0}{2}(Y+2)Y
X^2\nonumber\\
&&\hspace{-5mm}+\frac{\varepsilon^2\lambda_0^2}{2\omega^2}XY\left[\lambda_0
Y(1+Y)^2-\frac{\mu_0}{2}(2+12Y+18Y^2+5Y^3)\right]\nonumber\\
\label{effham}
\end{eqnarray}
which is time-independent. The first two terms come from the
unperturbed Hamiltonian, Eq.~(\ref{h0pert}), the other two terms
describe a rectified $\varepsilon^2$-correction coming from the high-frequency
perturbation. The mean-field fixed point and fluctuational fixed point of the effective Hamiltonian are
$$
A \left[q=\Omega
\left(1-\frac{\lambda^2\varepsilon^2}{2\omega^2}\right),\,p=0\right]\;\;\;\;\mbox{and}\;\;\;B(q=0,\,p=-1)\;,
$$
respectively. As the effective Hamiltonian is time-independent, there is no need
in the Poincar\'{e} section in this limit. Two ``trivial" zero-energy lines of the effective Hamiltonian are $X=0$ and $Y=0$. The non-trivial zero-energy line yields the effective instanton $X_{0}(Y)$.
In the
leading order of $\alpha\equiv \omega/\lambda_0\gg 1$ we obtain
\begin{equation}\label{optpatheff}
X_{0}(Y)\simeq\frac{2\Omega(1+Y)}{2+Y}\left[1-\frac{\varepsilon^2(2+10Y+15Y^2+4Y^3)}{2\alpha^2(2+Y)}\right]\,.
\end{equation}
The action along the effective instanton is
\begin{equation}\label{Kapitsaaction}
{\cal S}=-\int_{0}^{-1}X_{0}(Y)dY={\cal S}_0\left(1-
\frac{K\varepsilon^2}{\alpha^2}\right)\,,
\end{equation}
where
\begin{equation}\label{B}
K=\frac{6\ln 2-49/12}{1-\ln 2}= 0.2462\dots\,.
\end{equation}
We will call the resulting correction to the unperturbed action,
\begin{equation}\label{Kapitsadelta}
    \Delta {\cal S}_K= - K {\cal S}_0 (\varepsilon/\alpha)^2\,,
\end{equation}
the Kapitsa correction. The eikonal approximation, that has led to Eq.~(\ref{Kapitsadelta}),  demands $\Delta {\cal S}_K\gg 1$. Figure~\ref{highfreq} shows the ratio of $\Delta {\cal S}_K$ and the
correction to action, found by solving the Hamilton equations numerically (see
subsection \ref{numerics} for details of the numerics). As expected, good agreement is
observed at high frequencies, where the Kapitsa method is applicable.

\begin{figure}
\includegraphics[width=6.75cm,height=6.25cm,clip=]{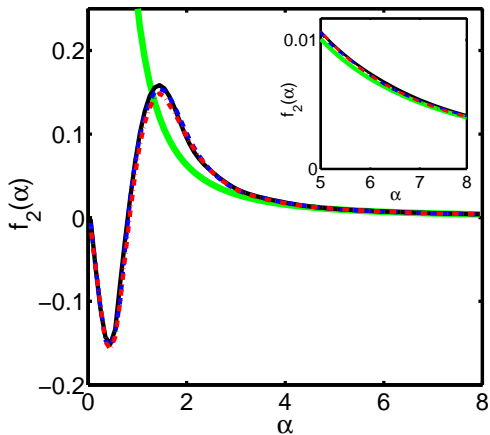}
\caption{(Color online.) Shown is the function $f_2(\alpha)$ determined numerically, see text. The curves for
$\varepsilon=0.04$ (solid line),
$\varepsilon=0.1$ (dashed line), and $\varepsilon=0.2$ (dash-dotted
line) collapse into a single curve as expected from Eq.~(\ref{delScomp}). The
$\alpha \gg 1$ asymptote of $f_2(\alpha)$, given by
Eq.~(\ref{f2eq}) (thick solid line), agrees with the numerical
result at high frequencies. The inset shows a blowup of the high-frequency region.} \label{f2}
\end{figure}

Now we can put the results obtained with the LT and with the
Kapitsa method into a broader context. The correction
to the unperturbed action $\Delta{\cal S}={\cal S}-{\cal S}_0<0$ must have the following general
form:
\begin{equation}\label{delScomp}
\frac{\Delta {\cal S}}{{\cal
S}_0}=-f_1\left(\alpha\right)|\varepsilon|-f_2\left(\alpha\right)\varepsilon^2+\dots\,.
\end{equation}
(Clearly, $\Delta{\cal S}$ must be analytical in $|\varepsilon|$, rather than in $\varepsilon$, as changing the sign of $\varepsilon$ only brings about a phase shift $\pi$ which changes the minimum point $t_0$ but leaves the minimum action unchanged.) The function $f_1(\alpha)>0$ is given by the LT, see Eq.~(\ref{minact}):
\begin{equation}\label{f1eq}
f_1(\alpha)=\frac{\pi\left\{\left[\sin(\alpha\ln
2)-\alpha\right]^2+\left[\cos(\alpha\ln
2)-1\right]^2\right\}^{1/2}}{(1-\ln 2)\sinh(\pi\alpha)}\,.
\end{equation}
The high-frequency asymptote of
$f_2(\alpha)$ is given by Eq.~(\ref{Kapitsaaction}):
\begin{equation}\label{f2eq}
f_2(\alpha)=\frac{K}{\alpha^2}\;\,,\;\;\;\;\alpha\gg
1\,.
\end{equation}
To calculate the function $f_2(\alpha)$ analytically for all
$\alpha$ would be quite cumbersome, so we found it numerically by
solving the Hamilton equations for three different values of
$\varepsilon$ and for different frequencies (see subsection
\ref{numerics} for details of the numerics). For each set of
$\varepsilon$ and $\alpha$ we computed $\Delta {\cal S}$ and used
Eq.~(\ref{delScomp}) with $f_1(\alpha)$ from Eq.~(\ref{f1eq}) to
extract $f_2(\alpha)$. As expected, the resulting plots of
$f_2(\alpha)$ for different $\varepsilon$ collapse into a single
curve, see Fig.~\ref{f2}. Noticeable is a non-monotonic,
alternating-sign $\alpha$-dependence. The high-frequency asymptote
of $f_2(\alpha)$ is in excellent agreement with the Kapitsa
correction (\ref{f2eq}). Note that $f_2(\alpha)$ vanishes as $\alpha
\to 0$. The reason is that, as $\omega \to 0$, the dependence of the
eikonal part of $\Delta {\cal S}_0$ on $\varepsilon$ becomes linear,
see Section \ref{adiabatic}.  This is a non-generic property of the
case when the branching rate $\lambda$ is modulated, while the
annihilation rate $\mu$ is not.

When does the Kapitsa correction to action dominate over the
exponentially small first-order correction at high frequencies? The
corresponding condition $\varepsilon^2 f_2(\alpha) \gg |\varepsilon|
f_1(\alpha)$ can be rewritten as $|\varepsilon|\gg \alpha^3
e^{-\pi\alpha}\,;$ it is easily satisfied for large $\alpha$. In
this regime the logarithmic susceptibility $\chi$ is proportional to
$|\varepsilon|$.

\subsection{Numerical calculations of the perturbed instanton and the action}\label{numerics}
\begin{figure}
\includegraphics[width=7cm,height=7cm,clip=]{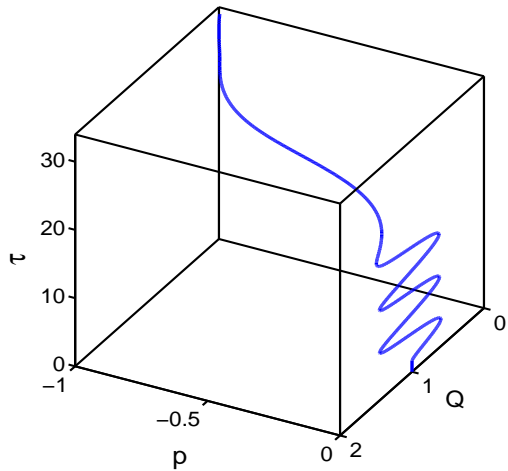}
\caption{(Color online.) A non-perturbative instanton trajectory obtained by solving
numerically the rescaled Hamilton equations (\ref{scaled}) for
$\varepsilon=0.6$ and $\alpha=1$. The trajectory first performs
large-amplitude oscillations around the mean-field fixed point $A$
of the unperturbed system [or, in other words, stays close to the
perturbed fixed point of the Poincar\'{e} map
$A_{\varepsilon}(t_0)$, then
leaves the vicinity of the fixed point and ultimately reaches (a
close vicinity of) the fluctuational point $B (Q=0,\,p=-1)$ which,
for the branching-annihilation reaction, remains unperturbed.}
\label{numS3d}
\end{figure}

\begin{figure}
\includegraphics[width=8.5cm,height=8cm,clip=]{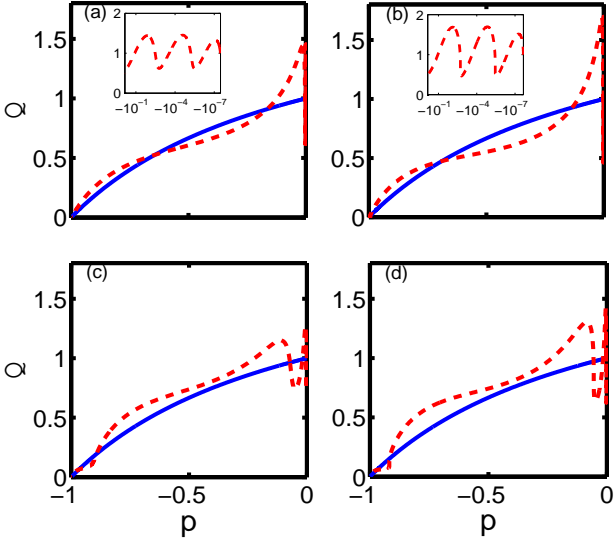}
\caption{(Color online.) Projections on the $(Q,p)$-plane of instantons found
numerically for four sets of parameters (the dashed lines). The
parameters and the corresponding rescaled actions [see
Eq.~(\ref{numaction})] ${\cal S}/\Omega$ are: (a) $\varepsilon=0.6$,
$\alpha=1$, ${\cal S}/\Omega=0.467$ (b) $\varepsilon=0.9$,
$\alpha=1$, ${\cal S}/\Omega=0.398$ (c) $\varepsilon=0.6$,
$\alpha=2$, ${\cal S}/\Omega=0.576$, and (d) $\varepsilon=0.9$,
$\alpha=2$, ${\cal S}/\Omega=0.545$. The solid line denotes the
unperturbed instanton (\ref{q(p)}). For comparison, the unperturbed
rescaled action is ${\cal S}_0/\Omega=2(1-\ln 2)=0.6137\dots$. The blowups in
the vicinity of $p=0$ (the insets) show that $Q(p)$ oscillates
before departing toward the fluctuational fixed point. In these examples $\delta=10^{-6}$.} \label{numS}
\end{figure}

To verify our analytical results, and to explore the parameter
regions beyond the validity of the perturbation methods, we computed the action numerically by solving the Hamilton
equations. Rescaling time $\tau=\lambda_0 t$ and the coordinate
$Q=\mu_0 q/\lambda_0=q/\Omega$ in
Eqs.~(\ref{pertHam})-(\ref{h1pert}), one obtains rescaled Hamilton equations
\begin{eqnarray}
\frac{\partial Q}{\partial \tau} &=& (1+2p)Q -(1+p)Q^2 + \varepsilon(1+2p)Q\cos \alpha \tau\,,\nonumber\\
\frac{\partial p}{\partial \tau} &=& -(1+p)p + (2+p)pQ -
\varepsilon(1+p)p\cos \alpha \tau\,,\label{scaled}
\end{eqnarray}
which follow from the rescaled Hamiltonian
$$
h(Q,p,\tau)=(p+1)pQ-\frac{1}{2}(2+p)p
Q^2+\varepsilon(p+1)p Q\cos \alpha \tau\,.
$$
The action along the perturbed instanton $[Q_{*}(t),p_{*}(t)]$ is given by
\begin{equation}\label{numaction}
{\cal S}=\Omega \int_{-\infty}^{\infty}\left\{-Q_{*}(\tau)\dot{p}_{*}(\tau)-h[Q_{*}(\tau),p_{*}(\tau),\tau]\right\}d\tau\,.
\end{equation}
where the integrand is fully determined by two dimensionless parameters:
the rescaled modulation amplitude $\varepsilon$ and rescaled frequency
$\alpha=\omega/\lambda$. The numerical solution is obtained by
shooting. We start at $\tau=0$ at a point $[Q(0),p(0)]$ lying on a small circle of radius $\delta \ll 1$
centered in the unperturbed
mean-field point $A(Q=1,p=0)$. The polar angle $\theta$ of the point $[Q(0),p(0)]$ on the circle serves
as the only shooting parameter: it is varied until the numerical trajectory
reaches, at some time $\tau=\tau_f$, (a close proximity of) the
fluctuational point $B(Q=-1,p=0)$. For the numerically found instanton $[Q_{*}(t),p_{*}(t)]$ to be a good approximation
to the true perturbed instanton in its entirety, $\delta$ must be sufficiently small.
On the other hand, too a small $\delta$ causes a long $\tau_f$ (and, therefore, a long
computation time and possible accumulation of numerical errors) because of
the intrinsic logarithmic slowdown near the fixed point.
By repeating the computations for a smaller circle,
$\delta_1<\delta$, we checked that
the same numerical instanton trajectory is reconstructed with high accuracy, apart from
additional oscillations that appear in the vicinity of the mean-field point of the unperturbed system (see Figs.~\ref{numS3d} and \ref{numS}). As long as $\delta$ is chosen to be sufficiently small, the contribution of these oscillations to the net action along the instanton is small.

We used this algorithm to verify our perturbative theoretical
results, see Figs. \ref{extvsw}-{\ref{f2}}. We also computed the
perturbed instanton numerically in the parameter region beyond the
validity of the perturbation techniques: $\varepsilon\lesssim 1$
and $\alpha\lesssim 1$. These computations strongly suggest that the
instanton connection persists for any $\varepsilon<1$ and any
$\alpha$. Figures~\ref{numS3d} and \ref{numS} show examples of
numerically found instantons, and action along them, for
non-perturbative values of parameters.

\section{Adiabatic approximation}\label{adiabatic}
Adiabatic approximation is non-perturbative in the modulation amplitude $\varepsilon$, and demands that the modulation frequency be much smaller than the characteristic relaxation rate of the system. Consider first a general single-species birth-death process with
\textit{time-independent} rates which possesses a non-trivial
quasi-stationary state and a single absorbing state at zero.
At times much longer than the time of relaxation
to the quasi-stationary state, the extinction probability at time $t$
can be represented as
$$
{\cal P}_0(t)=1- C e^{- r_{ex}^{(0)} t}\,,
$$
where $r_{ex}^0=r_{ex}^{(0)}(\lambda_1,\lambda_2,\dots)$ is the
extinction rate: the lowest excited eigenvalue $E_1$, see
\textit{e.g.} Refs. \cite{Assaf1,Assaf2}. The extinction rate
depends on the reaction rates of the problem
$\lambda_1,\lambda_2,\dots$. The constant $C$ is determined by the
initial conditions; for a sufficiently large initial number of
particles, $C$ is close to unity \cite{Assaf1,Assaf2}.

Now we introduce an adiabatically slow variation into the reaction
rates so that the characteristic time of the variation is much
longer than the relaxation time of the system, but much shorter than
the MTE. The extinction probability can now be written as
\begin{equation}\label{extprob10}
    {\cal P}_0(t)=1- C e^{- \int_0^t r_{ex}(t^{\prime}) \,dt^{\prime}}\,,
\end{equation}
where $r_{ex}(t)=r_{ex} [\lambda_1(t),\lambda_2(t),\dots]$ is the
instantaneous value of the slowly time-dependent extinction rate.
The average extinction rate $\bar{r}_{ex}$ can be defined via the
relation
\begin{equation}\label{extprob20}
{\cal P}_0(t)=1- C e^{- \bar{r}_{ex} t}\,.
\end{equation}
Comparing Eqs.~(\ref{extprob10}) and (\ref{extprob20}), we obtain
\begin{equation}\label{average10}
  \bar{r}_{ex}=\frac{1}{T} \int_0^T r_{ex}(t^{\prime}) \,dt^{\prime}\,,
\end{equation}
where $T$ is much longer than the relaxation time but much shorter than the MTE.
For slow periodic rate modulations with frequency $\omega$ Eq.~(\ref{average10}) can be rewritten as
\begin{equation}\label{mettime}
  \bar{r}_{ex}=\frac{2\pi}{\omega} \int_0^{2 \pi/\omega} r_{ex}(t^{\prime}) \,dt^{\prime}\,.
\end{equation}
The MTE is equal, in the adiabatic approximation, to $1/\bar{r}_{ex}$.

Now we illustrate the adiabatic approximation on our
branching-annihilation example. Here the periodic solution of the
mean-field equation (\ref{dotqmf}) is $q(t)\simeq \Omega
\left(1+\varepsilon \cos\omega t\right)$: the mean-field fixed point
follows the slowly changing rate $\lambda(t)$ adiabatically.

The extinction rate of the branching-annihilation system with time-independent
rates, was found, including the pre-exponent, in Refs. \cite{turner,kessler,Assaf1}:
\begin{equation}\label{extrate0}
r_{ex}^{(0)}=\frac{\Omega^{3/2}}{2\sqrt{\pi}}e^{-{\cal S}_0}\,,
\end{equation}
where ${\cal S}_0=2\Omega (1-\ln2)$ and $\Omega=\lambda/\mu \gg 1$.
We introduce a slow sinusoidal modulation in
the branching rate $\lambda$, see Eq.~(\ref{rates}), while keeping
$\mu=\mu_0$ constant. The instantaneous extinction rate becomes
\begin{equation}\label{instant}
    r_{ex}(t)=\frac{\Omega^{3/2}}{2\sqrt{\pi}} (1+\varepsilon\cos
\omega t)^{3/2}e^{-{\cal S}_0(1+\varepsilon\cos\omega t)}\,.
\end{equation}
For this result to be valid it is necessary that the argument in the
exponent be large: ${\cal S}_0(1+\varepsilon\cos\omega t) \gg 1$,
which yields
\begin{equation}\label{crit1}
1-|\varepsilon| \gg 1/\Omega\,.
\end{equation}
The instantaneous extinction rate (\ref{instant}) can be also
obtained, with exponential accuracy, in the following way. One
treats the time in the time-dependent eikonal Hamiltonian
(\ref{pertHam}) as a parameter and looks for zero-energy
trajectories as in the time-independent case.   In this way one
obtains the ``adiabatic instanton"
\begin{equation}\label{q(pt)}
q= q_a(p;t)=\frac{2\Omega(1+p)}{(2+p)}\,\left(1+\varepsilon \cos \omega t \right)\,,
\end{equation}
and finds the adiabatic action $S_0(t)=-\int_0^{-1}
q_a(p;t)\,dp={\cal S}_0 (1+\varepsilon \cos \omega t)$ which yields
(the minus of) the logarithm of the instantaneous extinction rate,
in agreement with Eq.~(\ref{instant}). This simple derivation yields
a necessary condition for validity of the adiabatic approximation.
Indeed, the adiabatic picture demands that $\omega$ be much less, at
all times, than the adiabatically varying relaxation rate
$\lambda(t)=\lambda_0 (1+\varepsilon \cos \omega t)$. Therefore, we
must require $\omega \ll \lambda_0(1-|\varepsilon|)$, that is
\begin{equation}\label{crit2}
\alpha \ll 1-|\varepsilon|
\end{equation}
For small or moderate modulation amplitudes, $|\varepsilon| \lesssim
1$, one recovers the same low-frequency criterion $\alpha \ll 1$
that led to Eq.~(\ref{minactadia}). As $|\varepsilon|$ closely
approaches 1, however, the criterion (\ref{crit2}) becomes much more
stringent. The criteria~(\ref{crit1}) and (\ref{crit2}) can be
rewritten as
\begin{equation}\label{critgen}
1-|\varepsilon| \gg \mbox{max}\, (\Omega^{-1}, \alpha)\,.
\end{equation}

\begin{figure}
\includegraphics[width=6cm,height=5.75cm,clip=]{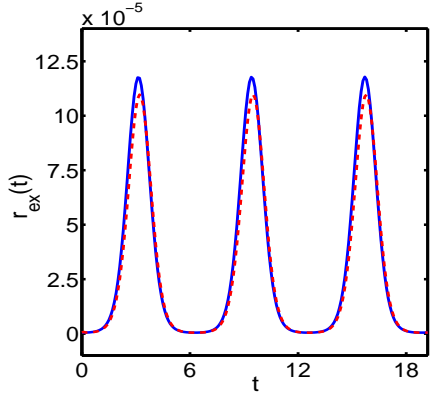}
\caption{(Color online.) Comparison of the adiabatic instantaneous extinction
rate, given by Eq.~(\ref{instant}) (the solid line) and the
instantaneous extinction rate determined from a numerical solution
of the master equation Eq.~(\ref{master}). The numerical extinction
rate (the dashed line) was computed as
$-(d/dt)\ln[1-{\cal
P}_0(t)]$. The parameters are $\Omega=25$, $\varepsilon=0.2$, and
$\alpha=0.04$.} \label{e1time}
\end{figure}

To verify Eq.~(\ref{instant}), we solved numerically (a truncated
version of) the original master equation (\ref{master}). The
numerical instantaneous extinction rate was computed from
$-(d/dt)\ln[1-{\cal P}_0(t)]$.  Figure~\ref{e1time} compares
Eq.~(\ref{instant}) with a numerical result for a small rescaled
frequency $\alpha=0.04$ and $\varepsilon=0.2$, and good agreement is
observed. The deviation in the peaks is mainly caused by
$1/\Omega$-corrections to the eikonal theory: we checked that it
goes down as $\Omega$ increases.

The average extinction rate (\ref{mettime}) can be written as
\begin{eqnarray}
\hspace{-1.5mm}\bar{r}_{ex}=\frac{\Omega^{3/2}}{4\pi^{3/2}}\int_{0}^{2\pi}(1+\varepsilon\cos
\tau)^{3/2}e^{-{\cal S}_0(1+\varepsilon\cos \tau)}d\tau\,. \label{extrate}
\end{eqnarray}
This expression is $\omega$-independent. For a comparison with the LT
prediction (see below), it will be convenient to compute
$-\ln(\bar{r}_{ex})$: the total action which includes, in addition to
the geometrical-optics contribution, the physical-optics
contribution coming from the pre-exponent.

\begin{figure}
\includegraphics[width=7.5cm,height=7cm,clip=]{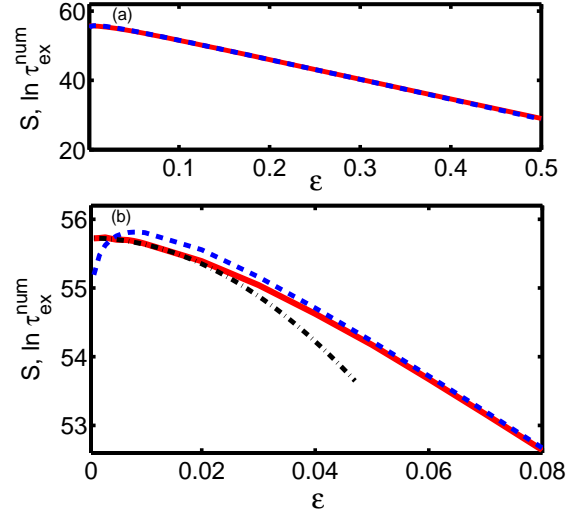}
\caption{(Color online.) (a) Comparison between the action ${\cal S}$ given by
Eq.~(\ref{specaction}) for $\varepsilon\Omega\gg 1$ (the dashed line) and the natural logarithm
of the MTE found by solving the master equation numerically (the
solid line), as a function of $\varepsilon$. (b) A blowup of the small-$\varepsilon$
region: shown are the $\varepsilon\Omega\ll 1$ asymptote
of the action ${\cal S}$ [Eq.~(\ref{logmet})] (the dash-dotted
line), the $\varepsilon\Omega\gg 1$ asymptote of the action ${\cal
S}$ [ Eq.~(\ref{specaction})] (the dashed line), and the natural
logarithm of the MTE found by solving the master equation
numerically (the solid line), as a function of $\varepsilon$. The
parameters in (a) and (b) are $\Omega=100$ and $\alpha=0.05$.
One can see that Eq.~(\ref{logmet}) breaks down when
$\varepsilon\Omega$ ceases to be small, whereas for
$\varepsilon\Omega \gg 1$ the $\varepsilon$-dependence becomes
approximately linear, and the action is well described by Eq.~(\ref{specaction}).}
\label{extvsepssmall}
\end{figure}

The integral in Eq.~(\ref{extrate}) can be calculated analytically in two
limits. For $|\varepsilon| \Omega\gg 1$ one can employ the saddle point
approximation and obtain
\begin{equation}\label{specadia}
\bar{r}_{ex}=\frac{\Omega (1-|\varepsilon|)^{3/2}}{4\pi \sqrt{|\varepsilon| (1-\ln
2)}}e^{-{\cal S}_0(1-|\varepsilon|)}\,.
\end{equation}
The saddle point is located at $t_*=\pm \pi/\omega$
(for $\varepsilon \gtrless 0$, respectively);
the
effective width of the Gaussian is
$\sigma\sim(2|\varepsilon|\Omega\omega^2)^{-1/2}\ll\pi/\omega$.  The
resulting action is
\begin{equation}
{\cal S}=-\ln(\bar{r}_{ex})={\cal
S}_0(1-|\varepsilon|)+\ln\left[\frac{\Omega (1-|\varepsilon|)^{3/2}}{4\pi \sqrt{|\varepsilon|(1-\ln
2)}}\right]. \label{specaction}
\end{equation}
The leading term ${\cal S}_0(1-|\varepsilon|)$ coincides with the
zero-frequency limit predicted by the LT of Section \ref{theory}:
see Eq.~(\ref{minactadia}) with the $\alpha^2$-term neglected. Its
physical meaning is transparent: in view of the adiabatically slow
rate modulation, the effective ``activation barrier" to extinction
${\cal S}_0(1-|\varepsilon|)$ is determined by the minimal value of
$\lambda(t)=\lambda_0 (1+\varepsilon \cos \omega t)$ which is equal
to $\lambda_0 (1-|\varepsilon|)$. Equation~(\ref{specaction}),
however, also includes an important pre-exponent (recall that
$\Omega \gg 1$), missed by the geometrical-optics order of the
eikonal expansion.  Note that, because of the additional
pre-exponent, coming from the saddle-point integration, the
$\Omega$-dependence of the resulting pre-exponent changes and
becomes $\sim \Omega$ instead of $\sim \Omega^{3/2}$.
Fig.~\ref{extvsepssmall}a compares the prediction of
Eq.~(\ref{specaction}) with the natural logarithm of the MTE
obtained by numerically solving the master equation.

In the opposite limit, $|\varepsilon| \Omega\ll 1$, the integral
(\ref{extrate}) can be calculated via a Taylor expansion of the integrand
in $\varepsilon\Omega$. Terms proportional to $\cos \tau$
vanish after the integration, and
the leading-order result is
\begin{equation}
\bar{r}_{ex}\simeq r_{ex}^{(0)}\left[1+\varepsilon^2\Omega^2(1-\ln 2)^2\right]\,.
\label{resextratesmall}
\end{equation}
This yields
\begin{equation}\label{logmet}
{\cal S}={\cal S}_0-\varepsilon^2\Omega^2(1-\ln 2)^2\,.
\end{equation}

Note that the strong inequality $|\varepsilon| \Omega \gg 1$ in
Eq.~(\ref{specaction}) coincides with the validity criterion of the
eikonal method in the linear theory (LT): the left inequality in
Eq.~(\ref{double}). Correspondingly, when the opposite strong
inequality $|\varepsilon| \Omega\ll 1$ holds, the correction to
${\cal S}_0$ that appears in Eq.~(\ref{logmet}) is much smaller than
$1$ and therefore strongly non-eikonal. Fig.~\ref{extvsepssmall}b
shows a comparison between Eq.~(\ref{logmet}) and the numerical
solution of the master equation, and excellent agreement is
observed.

One can notice that the leading-order term in the action
(\ref{specaction}), ${\cal S}_0 (1-|\varepsilon|)$, goes down
linearly in $|\varepsilon|$ for all $|\varepsilon|<1$. The linearity
in $|\varepsilon|$ is non-generic and stems from the fact that
$\lambda(t)/\mu_0$ is linear in $\varepsilon$, and so is the
exponent in the instantaneous extinction rate (\ref{extrate}). In
the general case, the $|\varepsilon|$-dependence of the exponent
that appears in the instantaneous extinction rate is nonlinear, and
therefore the linear $|\varepsilon|$-dependence of the action (which
also appears in the LT-regime of the eikonal approximation) will
break down at $|\varepsilon| ={\cal O}(1)$. This is indeed what one
observes if, instead of modulating $\lambda$, he modulates $\mu$:
\begin{equation}\label{rates2}
\lambda(t)=\lambda_0\,,\;\;\;\mu(t)=\mu_0\left(1+\varepsilon \cos
\omega t\right)\,,
\end{equation}
In this case, using Eq.~(\ref{mettime}), we obtain the following
action for $|\varepsilon|\Omega\gg 1$:
\begin{equation}
{\cal S}=-\ln(\bar{r}_{ex})=\frac{{\cal
S}_0}{1+|\varepsilon|}+\ln\left[\frac{\Omega}{4\pi \sqrt{|\varepsilon|(1+|\varepsilon|)(1-\ln
2)}}\right]. \label{specactionmu}
\end{equation}
For $|\varepsilon|\ll 1$ the leading-order term ${\cal
S}_0/(1+|\varepsilon|) \simeq {\cal S}_0 (1-|\varepsilon|)$
coincides with the leading term of Eq.~(\ref{specaction}), or with
Eq.~(\ref{minactadia}) where one must put $\alpha=0$. For not small
$|\varepsilon|$, however, the $|\varepsilon|$-dependence is
non-linear.

\section{Summary and Discussion}\label{summary}

\begin{figure}
\includegraphics[width=8.0cm,clip=]{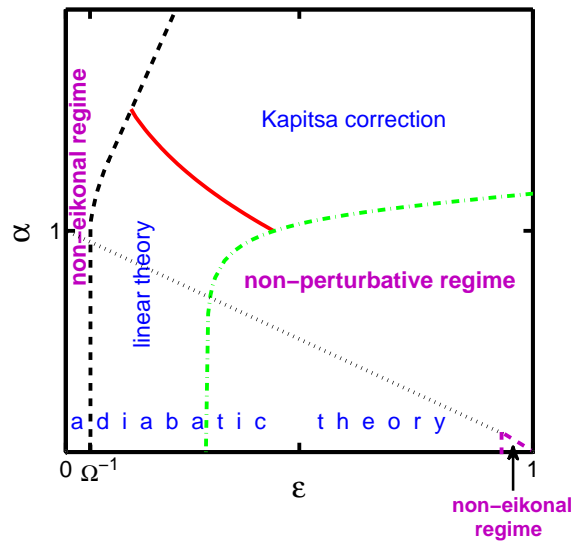}
\caption{(Color online.) A schematic diagram of different regimes of
extinction on the parameter plane  $(\varepsilon>0, \alpha)$ of the
branching-annihilation process. Either the LT, or the Kapitsa method
is applicable well to the left of and above the dash-dotted line
(where the inequality $\Delta{\cal S}/{\cal S}_0\ll 1$ holds) and
well to the right of the dashed line (where $\Delta{\cal S}\gg 1$).
On the solid line  $\varepsilon=f_1(\alpha)/f_2(\alpha)$, see Eqs.
(\ref{delScomp})-(\ref{f2eq}). The adiabatic approximation holds
well below the dotted line. The horizontal size of the non-eikonal
regions at $\alpha \to 0$ is of order $1/\Omega$.  The rest of the
diagram is self-explanatory.} \label{paramphase}
\end{figure}

We have developed three complementary perturbation techniques for an
analytical calculation of the exponential reduction in the mean time
to extinction (MTE) in single-species birth-death processes (or
reaction kinetics) that occur in a time-periodic environment. These
are the linear theory (valid at small modulation amplitudes), the
Kapitsa method (valid at high modulation frequencies), and the
adiabatic approximation (valid at low modulation frequencies). We
presented our theory on the example of a simple
branching-annihilation process. Figure~\ref{paramphase} shows
different regimes of this process on the parameter plane
$(\varepsilon>0, \alpha)$.

The linear theory and the Kapitsa method, as we used them, are
rooted in the time-dependent eikonal theory that employs a large
parameter $\Omega$: the average number of individuals in the
quasi-stationary state at times short compared with the MTE. The
adiabatic approximation also employs the large parameter $\Omega$.
It yields, however, a more accurate result than the other two
techniques as it yields a nontrivial $\Omega$-dependent pre-exponent
of the MTE. The higher accuracy, achieved in this way, is important
in view of the fact that, at low frequencies, the reduction in the
MTE is the largest.

An important result of this work is that, at high modulation
frequencies and not too small modulation amplitudes, the linear
theory can greatly underestimate the true reduction in the MTE. In
this regime an accurate result is given by the Kapitsa correction,
calculated in this work.

Our calculations of the modulation-induced reduction in the MTE are
closely related to finding the optimal path to extinction, or
instanton connection \cite{dykman1,Kamenev1,escudero}. Although in
generic Hamiltonians the existence of an instanton connection (a
more standard mathematical term is a heteroclinic trajectory) can be
guaranteed only at very small modulation amplitudes
\cite{guckenheimer}, our numerical results strongly indicate that,
in the birth-death processes with an absorbing state at zero, a
perturbed instanton persists at \textit{any} reasonable modulation
amplitude and at any frequency.  This remarkable fact must be
intimately related to the non-generic structure of the birth-death
Hamiltonians. More specifically, a future proof of the instanton
persistence at arbitrary (physically reasonable) time-dependent
perturbations of the reaction rates will most likely exploit the
invariance of the zero-energy mean-field line $p=0$ and the
zero-energy extinction line $q=0$ in the perturbed Hamiltonian.

\section*{Acknowledgments} We are very grateful to Mark Dykman for
enlightening discussions. M.~A. was supported by the Clore
Foundation. M.~A. and B.~M. were supported by the Israel Science
Foundation (grant No. 408/08); A.~K.  was supported by the NSF grant
DMR-0405212  and by the A.~P.~Sloan foundation. M.~A. and B.~M. are
grateful to the William I. Fine Theoretical Physics Institute of the
University of Minnesota for hospitality.


\begin{thebibliography}{99}
\bibitem{badeg} Invariance of the fluctuational fixed point under rate modulations is a non-generic feature of the branching-annihilation model. In general (for instance, for the more general model $A\to 2A$, $2A\to 0$ and $A\to 0$ \cite{AKM,Kamenev2}) both the mean-field point, and the
fluctuational point are perturbed.
\bibitem{escudero1} Escudero and Rodr\'{\i}guez \cite{escudero} showed, for
a general (not necessarily sinusoidal) time-periodic modulation of the reaction rates $\lambda(t)$ and $\mu(t)$ that
$M(t_0)$ is time periodic with zero mean. Therefore, there
exists some $t_0^*$ for which $M(t_0^*)=0$.
\end{thebibliography}
\end{document}